\documentclass[aps, prl, twocolumn, a4paper, superscriptaddress]{revtex4-2}
\usepackage{graphicx}
\usepackage[paperwidth=215mm,paperheight=300mm,centering,hmargin=1.6cm,vmargin=2cm]{geometry}
\usepackage{dcolumn}
\usepackage{bm}
\newcommand*{\citenst}[2][]{%
  \begingroup
  \let\NAT@mbox=\mbox
  \let\@cite\NAT@citenum
  \let\NAT@space\NAT@spacechar
  \let\NAT@super@kern\relax
  \renewcommand\NAT@open{[}%
  \renewcommand\NAT@close{]}%
  \citep{#2}%
  \endgroup
}
\bibpunct{[}{]}{,}{n}{}{,}
\usepackage{graphicx, amsmath, amssymb}
\usepackage{siunitx}
\usepackage{hyperref}
\usepackage{float}
\usepackage{afterpage}
\newcommand{\rom}[1]{\uppercase\expandafter{\romannumeral #1\relax}}
\usepackage{lipsum}

\begin{document}

\preprint{APS/123-QED}

\title{Waveguide and cavity quantum electrodynamics with topological bowtie modes}

\author{Anastasiia V.~Vladimirova}
\affiliation{Department of Electrical and Photonics Engineering, DTU Electro, Technical University of Denmark, Building 343, DK-2800 Kgs.~Lyngby, Denmark.}
\affiliation{NanoPhoton - Center for Nanophotonics, Technical University of Denmark, Ørsteds Plads 345A, DK-2800 Kgs.~Lyngby, Denmark.}
\author{Guillermo Arregui}
\affiliation{Department of Electrical and Photonics Engineering, DTU Electro, Technical University of Denmark, Building 343, DK-2800 Kgs.~Lyngby, Denmark.}
\author{Sergei Lepeshov}
\affiliation{Department of Electrical and Photonics Engineering, DTU Electro, Technical University of Denmark, Building 343, DK-2800 Kgs.~Lyngby, Denmark.}
\affiliation{NanoPhoton - Center for Nanophotonics, Technical University of Denmark, Ørsteds Plads 345A, DK-2800 Kgs.~Lyngby, Denmark.}
\author{Christian Anker Rosiek}
\affiliation{Department of Electrical and Photonics Engineering, DTU Electro, Technical University of Denmark, Building 343, DK-2800 Kgs.~Lyngby, Denmark.}
\author{Babak Vosoughi Lahijani}
\affiliation{Department of Electrical and Photonics Engineering, DTU Electro, Technical University of Denmark, Building 343, DK-2800 Kgs.~Lyngby, Denmark.}
\affiliation{NanoPhoton - Center for Nanophotonics, Technical University of Denmark, Ørsteds Plads 345A, DK-2800 Kgs.~Lyngby, Denmark.}
\author{Søren Stobbe}
\affiliation{Department of Electrical and Photonics Engineering, DTU Electro, Technical University of Denmark, Building 343, DK-2800 Kgs.~Lyngby, Denmark.}
\affiliation{NanoPhoton - Center for Nanophotonics, Technical University of Denmark, Ørsteds Plads 345A, DK-2800 Kgs.~Lyngby, Denmark.}

\date{\today}

\begin{abstract}
We present a theoretical study on photonic topological crystals whose symmetry is governed by quantum valley-Hall topological insulators and whose propagating edge modes are strongly confined due to bowtie geometries.
Dielectric bowtie structures exploit the field discontinuities at boundaries between materials with different refractive indices, and here bowties emerge at the topological interface due to the close proximity of two triangular features in the underlying crystal.
The topological bowtie mode features a unit-cell mode volume down to $8\times10^{-4}$ cubic wavelengths at the center of the bowtie bridge of width $\SI{10}{\nm}$, and we show that it is possible to use perturbed versions of the unit cells as building blocks for topological heterostructure bowtie cavities with quality factors exceeding $10^7$. Due to the tightly confined bowtie mode, this implies a strongly enhanced light-matter interaction as quantified by a Purcell factor of $3 \times 10^6$.
\end{abstract}

\maketitle

\section{\label{sec:intro}Introduction}

Controlling the interaction between light and matter is fundamental to scientific and technological developments in sensing and imaging~\cite{zhen2013enabling, altug2022advances, beliaev2022pedestal}, quantum optics and quantum information processing~\cite{RevModPhys.87.347}, and nanoscale light sources~\cite{romeira2018purcell}. To this end, photonic crystals have proven to be a highly versatile platform because they strongly modify the local density of optical states (LDOS)~\cite{RevModPhys.87.347}. By varying the constitutive material, lattice parameters, and structural symmetries, photonic crystals can be used to engineer the dispersion of light~\cite{lopez2003materials}, to either enhance or suppress light-matter interaction, and more generally to realize fundamental components for experiments on quantum electrodynamics, such as cavities and waveguides.

Photonic topological insulators (PTIs) are a particular class of photonic crystals, which have recently attracted significant attention~\cite{ozawa_topological_2019, kim2020recent, ni2023topological}. In dielectric PTIs, which are governed by time-reversal symmetry, topological edge modes may be created by exploiting the valley degree of freedom~\cite{Ma_2016}, which results in a photonic analog of the quantum valley-Hall (QVH) effect. Although the origin of QVH interface states and conventional line-defect waveguide modes differ, they both exhibit propagating modes~\cite{ozawa_topological_2019, kim2020recent}, slow light~\cite{yoshimi2020slow} and chiral light-matter interaction~\cite{lodahl2017chiral, hauff2022chiral}. Besides, both are predicted to be prone to strong backscattering induced by fabrication imperfections~\cite{Bravo_2021}. Two recent experiments found clear signatures of Anderson localization~\cite{rosiek_observation_2022, arora2023multiple} but reached different conclusions on whether or not PTIs may offer a small advantage in terms of reduced backscattering.

\begin{figure}[b]
\includegraphics[width=1\linewidth] {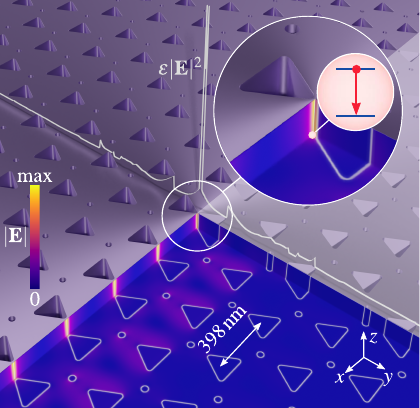}
\caption{\textbf{Strong light-matter interaction in dielectric topological bowtie waveguides}. The interface between two QVH PTIs rotated $180^{\circ}$ relative to each other creates a topological interface with opposing tips that form bowties at the interface. The structure features three symmetry planes, and $1/8$ of the structure has been cut out. A cutaway shows the electric field amplitude, $|\mathbf{E}|$, on cross-sections along the three symmetry planes of the structure. The normalized electromagnetic energy (white curve), $\varepsilon|\mathbf{E}|^2$, in the center of the structure and perpendicular to the waveguide axis shows a strongly enhanced intensity in the bowties, which greatly enhances the interaction with integrated quantum emitters (zoom-in).}
\label{fig:1}
\end{figure}

\begin{figure*}[htbp!]
\centering
\includegraphics[width=1\linewidth] {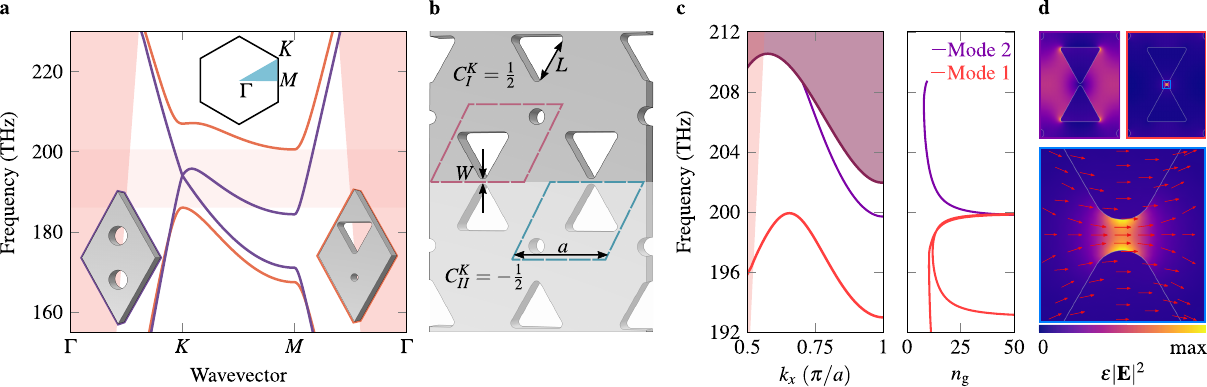}
\caption{
\textbf{Design of the topological bowtie waveguide.}
(a) Band structure of the bulk crystal before (purple) and after (orange) breaking of inversion symmetry by reducing the radius, $R: \SI{50}{nm} \rightarrow \SI{20}{nm}$, of one circular hole and replacing the other with an equilateral triangle with the rounded corners and the sidelength $L = \SI{173}{nm}$, measured from end to end of the radius of each rounded corner. The degeneracy lifting at the $K$ point manifests the topological transition between energy levels. Red-shaded areas indicate the light cone and the opened gap. The top and bottom insets show the irreducible Brillouin zone and the unit cells geometries, respectively.  
(b) Interface waveguide geometry with lattice constant $a=\SI{398}{nm}$. The unit cell of the top crystal (pink dashed line) is related to the unit cell of the bottom crystal (blue dashed line) by the mirror reflection. The triangle sidelength, $L$, the void radius of curvature at the triangle corners, $r = \SI{10}{nm}$, and $a$, determine the bridge width, $W = \SI{10}{nm}$.
(c) Dispersion diagram and group index for the waveguide modes.
(d) Energy density at the edge of the Brillouin zone $(k_x = \pi/a)$ for the waveguide modes. The zoom-in shows the hot-spot formation in the bowtie bridges for the lower-frequency mode. The arrows show the local field polarization.}
\label{fig:2}
\end{figure*}

The LDOS may also be enhanced without periodic structures but solely relying on the electromagnetic boundary conditions across material interfaces. In particular, bowtie geometries have been predicted to enhance both linear and nonlinear optical interactions~\cite{gondarenko2006spontaneous, hu2016design, choi2017self, wang2018maximizing, christiansen2023impact} due to their ability to localize the optical energy in a very small volume. This confinement regime was recently demonstrated experimentally and constitutes a new avenue of nanocavity research, in part because such tight confinement of light was previously believed to be inaccessible without plasmonic effects and in part because the roles of disorder and critical dimension are very different compared to conventional dielectric cavities~\cite{albrechtsen2022nanometer, babar2023self}. Despite the recent progress, the physics and applications of bowtie cavities remains almost entirely unexplored. To date, no experiments have demonstrated modified light-matter interaction in bowtie cavities, and theoretical works have so far not considered the implications of bowtie geometries on waveguide quantum electrodynamics, let alone topological photonics.

In the present work, we explore the advantages of both aforementioned approaches and propose a QVH waveguide design with bowtie features at the interface. The proposed platform enhances light-matter interaction in topological states of light by forming optical hot-spots at dielectric bowtie bridges, as shown in Fig.~\ref{fig:1}, and, at the same time, by slowing down light. We evaluate the potential of this structure for applications in quantum nanophotonics by calculating the Purcell factor~\cite{purcell1995spontaneous}, and we show that the topological waveguide mode can be modified to form an adiabatic heterostructure cavity with a theoretical quality factor of $Q \sim 10^7$ and a mode volume of $V_{\text{mod}} = 6\times10^{-3}\lambda_{\text{res}}^3$.

\vspace{-1.125 cm} 
\begin{center}
\begin{figure*}[htbp!]
\includegraphics[width=1\linewidth] {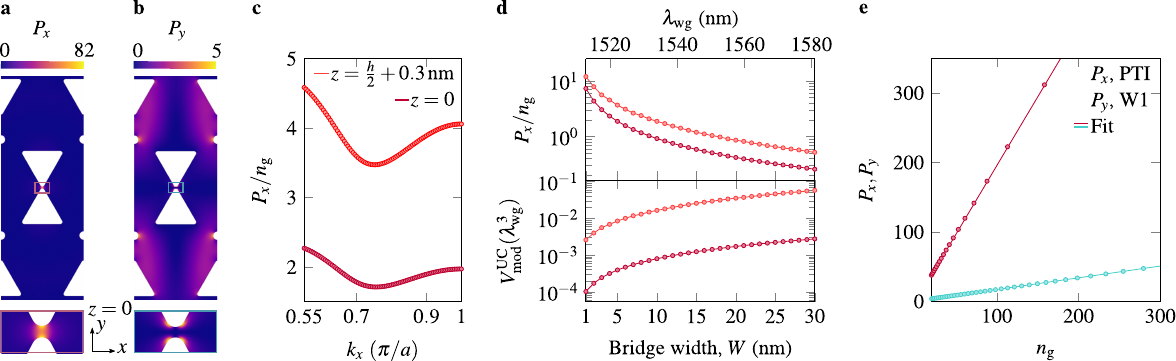}
\caption{\textbf{Purcell factor and mode volume}. (a) Spatial dependence of the Purcell enhancement for a point-source emitter with dipole moment along the $x$-axis. The evaluation is done at a fixed $n_\text{g} = 30$ using the geometry in Fig.~\ref{fig:2}(b). (b) Same as (a) for a dipole along the $y$-axis. (c) Evaluation of $P_x/n_\text{g}$ along the BZ for an emitter at the center of the bowtie (dark red) and in the air at a height $\SI{0.3}{nm}$ above the slab on top of the dielectric bridge center (light red). (d) Purcell factor $P_x/n_\text{g}$, and mode volume, $V_{\text{mod}}^{\text{UC}}$, for a fixed $n_\text{g} = 30$ and as a function of the bowtie width, $W$, which is modified by changing the triangle side-lengths everywhere. This causes the guided-mode wavelength, $\lambda_{\text{wg}}$, to shift as indicated by the top axis. Dark and light red curves represent the same as in (c). (e) Purcell factor as a function of the group index at $k_x\in[0.75, 1]$. The blue color corresponds to the W1 waveguide with a lattice constant $a=\SI{346}{\nm}$, and a hole radius $r=0.3\times a$ maintaining the same slab thickness as utilized in the modeling of PTI. The numerical simulation results are depicted by dots, while the line represents the linear fit to the data.}
\label{fig:3}
\end{figure*}
\end{center}

We aim to establish building blocks for PTI bowtie structures realized in dielectric slabs of height $h =\SI{220}{nm}$ and refractive index $n=3.48$, which corresponds to silicon in the telecom C-band. We consider honeycomb crystals with a lattice constant of $a = \SI{398}{nm}$ and unit cells (UCs) as shown in Fig.~\ref{fig:2}(a). For the UC with two circles, the crystal has $C_{6\text{v}}$ symmetry, and the band structure exhibits a Dirac-cone degeneracy at the $K$ point in the Brillouin zone (BZ). It is well known~\cite{Ma_2016, VH_review_2018, ozawa_topological_2019} that this photonic graphene turns into a time-reversal-invariant PTI upon reducing the symmetry to $C_{3\text{v}}$. We adhere to this procedure by replacing one of the circular holes with an equilateral triangular hole with rounded corners. The rounding is necessary to avoid lightning-rod effects leading to non-physical mode volumes~\cite{albrechtsen2022two}. Contrary to previous QVH waveguide realizations that employed triangles with edges parallel to the UC boundaries~\cite{shalaev2019robust, yoshimi2020slow}, we consider triangles whose corners face the UC boundaries. This particular UC geometry implies that a zigzag interface~\cite{ozawa_topological_2019} between one such topological crystal (denoted $\rom{1}$) and the mirror-symmetric version of it (denoted $\rom{2}$) results in a waveguide UC with a bowtie geometry at the center, as shown in Fig.~\ref{fig:2}(b). The length of the straight segment, $L$, and the radius of curvature, $r$, define the bowtie width, $W$. The radius of the complementary circular hole, $R$, serves as an additional degree of freedom that allows tailoring the dispersion relation of the waveguide mode. 

The proposed waveguide supports two guided modes, both exhibiting diverging group indices, $n_\text{g}$, at the BZ edge, facilitating a large LDOS~\cite{RevModPhys.87.347}, as shown in Fig.~\ref{fig:2}(c). The energy density of the two modes in the slab mid-plane ($z=0$) for $k_x = \pi/a$ is shown in Fig.~\ref{fig:2}(d). The low-frequency mode, Mode $1$, has an antinode in the bowtie bridge, and its local polarization is along the $x$-axis, i.e., such that the high-index bowtie acts as a field concentrator~\cite{albrechtsen2022two}. Mode $2$ does not exhibit highly confined energy in the bowtie bridges, and we will henceforth concentrate solely on Mode $1$.

To evaluate the strength of the light-matter interaction, we calculate the ratio of the spontaneous-emission decay rate in the QVH waveguide to that in the corresponding homogeneous medium (bulk silicon), i.e., the Purcell factor~\cite{purcell1995spontaneous}. In the dipole approximation, the Purcell factor for a point-like quantum emitter at a position $\mathbf{r}_\text{d}$, oriented along the unit vector $\mathbf{\hat{n}}_\text{d}$, and emitting at the waveguide mode frequency, $\omega$, can be derived analytically~\cite{Purcell-Hughes-2004},
\begin{equation}\label{eq: Purcell}P_{\mathbf{\mathbf{\hat{n}}_{\text{d}}}}(\mathbf{r}_{\text{d}},\omega) = \frac{3\pi c^2a n_{\text{g}}}{n(\mathbf{r}_{\text{d}})\omega^2}\frac{|\mathbf{E}(\mathbf{r}_{\text{d}})\cdot\mathbf{\hat{n}}_{\text{d}}|^2}{V_{\text{mod}}^{\text{UC}}(\mathbf{r}_{\text{d}})\varepsilon(\mathbf{r}_{\text{d}})|\mathbf{E}(\mathbf{r}_{\text{d}})|^2}.
\end{equation}
Here, $\mathbf{E}(\mathbf{r}_{\text{d}})$ and $V_{\text{mod}}^{\text{UC}}(\mathbf{r})$ are the electric field and the mode volume of the Bloch mode, respectively. The latter is given by~\cite{rao2007single} 
\begin{equation}\label{eq:V_mod}
V_{\text{mod}}^{\text{UC}}(\mathbf{r}) = \frac{\int_{\Omega_{\text{UC}}}d\mathbf{r'}\varepsilon(\mathbf{r'})\mathbf{E}(\mathbf{r'})\cdot\mathbf{E}^*(\mathbf{r'})}{\varepsilon(\mathbf{r})\mathbf{E}(\mathbf{r})\cdot\mathbf{E}^{*}(\mathbf{r})},
 \end{equation}
where $\Omega_{\text{UC}}$ is the volume of the waveguide UC. Using Eq.~(\ref{eq: Purcell}), we analyze the spatial dependence of the Purcell factors, $P_{x}$ and $P_{y}$, corresponding to $x$- and $y$-polarized dipole emitters, respectively. This is shown in Fig.~\ref{fig:3}(a) and (b) for emitters at the mid-plane of the slab ($z=0$) and at a fixed point in the dispersion relation for which $n_\text{g} = 30$. We consider this value of $n_\text{g}$ in order to limit the discussion to a regime where backscattering losses do not become prohibitively high for practically relevant device lengths~\cite{rosiek_observation_2022, arregui2023cavity}. We find a Purcell factor for $x$-polarized emitters placed in the center of the bowtie of $55$. Also, high Purcell factors up to $82$ are predicted for emitters located near the air tips (see the zoom-in in Fig.~\ref{fig:3}(a)), although this may likely be of limited experimental significance due to the immediate proximity of surfaces and their associated roughness leading to significant experimental and numerical uncertainties~\cite{garcia2017physics,albrechtsen2022two, hauff2022chiral}. 

We now consider how the Purcell factor varies in momentum space. To separate the well-known slow-light effect on the Purcell factor from the spatial confinement of interest here, we evaluate $P_{x}(\mathbf{r}_{\text{c}}, \omega)/n_{\text{g}}$. We consider the dependence on the wavevector, $k_x$, for two different points located at the center of the bowtie at the heights $z = 0$ and $z =  h/2 + \SI{0.3}{nm}$ as shown in Fig.~\ref{fig:3}(c). These two planes correspond to the decay-rate enhancement for emitters embedded in the center of the membrane, e.g., erbium ions~\cite{polman1997erbium, weiss2021erbium, gritsch2022narrow} and confined excitons in monolayer $2$D-materials deposited above the patterned slab~\cite{denning2022quantum}. Notably, $P_{x}(\mathbf{r}_{\text{c}}, \omega)/n_{\text{g}}$ for our design exhibits only a weak dependence on $k_x$. 

The Purcell enhancement can be substantially improved by reducing the width of the bowtie bridge~\cite{albrechtsen2022two}. Figures~\ref{fig:3}(d) and (e) illustrate the group-index-normalized Purcell factor and mode volume as a function of the width, $W$, for a fixed $n_{\text{g}} = 30$. We vary the bowtie width from $\SI{1}{nm}$ to $\SI{30}{nm}$ to illustrate the span between extreme bowtie structures, potentially enabled by self-assembly~\cite{babar2023self}, and bowties achievable in several semiconductor platforms. Such geometrical modification allows covering the full telecom C-band as shown in the top $x$-axis of Fig.~\ref{fig:3}(d, e). We observe that $P_x/n_{\text{g}}$ drops with increasing $W$, which results from the associated increase in $V_{\text{mod}}^{\text{UC}}$. For state-of-the-art silicon top-down nanofabrication, i.e., $W = \SI{8}{nm}$~\cite{albrechtsen2022nanometer}, we find $V_{\text{mod}}^{\text{UC}}|_{z=0} = 0.63\times10^{-3} (\lambda_{\text{wg}}^3)$ and $V_{\text{mod}}^{\text{UC}}|_{z=h/2+\SI{0.3}{nm}} = 13 \times10^{-3} (\lambda_{\text{wg}}^3)$. This leads to high Purcell factors up to $P_x^{z=0} = 70$ and $P_x^{z=h/2+\SI{0.3}{nm}} = 143$. Remarkably, these substantial enhancement factors are attained in a propagating waveguide mode, i.e., without a cavity. 

In order to investigate the benefits of the topological bowtie waveguides, we now compare their performance to conventional W1 waveguides, which have been studied extensively for waveguide quantum electrodynamics~\cite{RevModPhys.87.347}. Our waveguides and the W1 waveguides can be directly compared as they can be realized in the same materials at the same wavelengths, only it is important to note that the Purcell factor in the case of W1 is larger for $P_y$-oriented dipoles, so this is what we consider in Fig.~\ref{fig:3}(f), where we assume a realistic value of $W=\SI{10}{\nm}$. Remarkably, while the slope of $P_y$ is $0.2$ for the W1 waveguide, the slope of $P_x$ is $2.0$ for the topological bowtie waveguide. This twenty-fold enhancement relative to W1 waveguides at the same group index is a direct consequence of the bowtie field enhancement. In comparison to conventional W1 waveguides, which achieve high Purcell factors only due to the slow light effect, our structures additionally take advantage of the bowtie field confinement, spanning across the whole BZ, cf.\ Fig.~\ref{fig:3}(c).

\begin{center}
\begin{figure*}[hbtp!]
\includegraphics[width=1\linewidth] {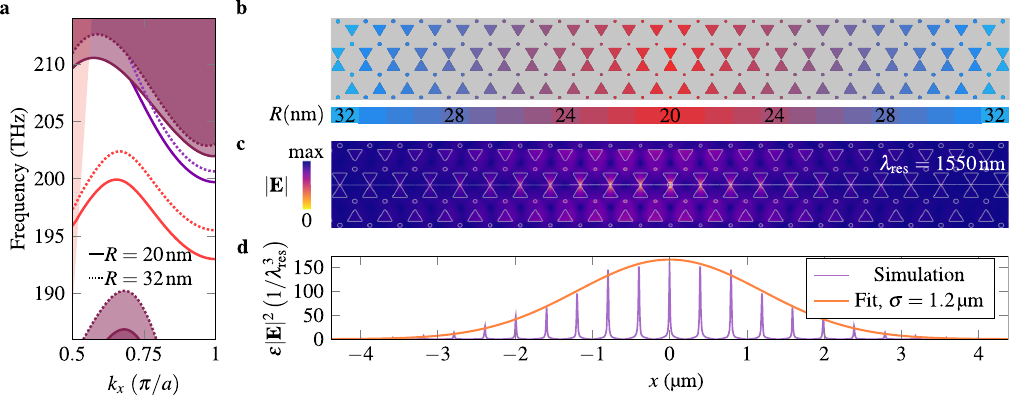}
\caption{\textbf{Heterostructure bowtie crystal cavity} (a) Band diagram of the central waveguide UC with $R = \SI{20}{nm}$ (solid lines) plotted on top of the band diagram of the mirror waveguide UC with $R = \SI{32}{nm}$ (dashed lines). (b) A sketch of the cavity design. The different radii of the circles in adjacent waveguide UCs are shown with different colors spanning from $R = \SI{20}{nm}$ to $R = \SI{32}{nm}$ with a $\SI{1}{nm}$ step.
(c) Electric field amplitude distribution of the fundamental mode of the cavity in the center of the slab ($z = 0$) obtained from a $3$D simulation. (d) Energy density distribution along the waveguide interface and the Gaussian envelop fitting the peaks. In subplots (b), (c), and (d), the $x$-coordinate spans a consistent range.}
\label{fig:4}
\end{figure*}
\end{center}
\vspace{-1.125 cm}
Finally, we utilize the topological waveguide mode to engineer a mode-gap cavity~\cite{akahane2003high}. Previous studies of optical cavities based on topological structures have focused on trapped topological $0$D states at the interface of 1D PTI structures~\cite{ota2018topological}, higher-order topological corner states between adjacent $2$D PTIs~\cite{ota2019photonic,zhang2020low, pti_bt_2022}, or topological ring cavities formed through 1D topological edge states that exhibit robust transport across sharp bends~\cite{smirnova2020room, noh2020experimental, zeng2020electrically, mehrabad2020chiral, gu2021topological, gong2020topological, liu2022topological}. Here, we use the bowtie QVH waveguide as a building block to form an ultra-high-$Q$ multi-step heterostructure cavity following the approach proposed in Ref.~\onlinecite{song2005ultra}. We introduce tapered mirrors by varying the geometrical parameters of the waveguide along the propagation direction such that the waveguide mode frequency at the BZ edge forms a smooth confinement potential. In our case, the cavity is created by changing $R$ from $\SI{20}{nm}$ in the cavity center to $\SI{32}{nm}$ at the cavity edges in steps of $\SI{1}{nm}$, resulting in tapered mirrors distributed along $N = 12$ UCs. The dispersion diagrams of the UCs with $R = \SI{20}{nm}$ and $R = \SI{32}{nm}$ are shown in Fig.~\ref{fig:4}(a), evidencing the creation of a $\SI{2.7}{THz}$-deep potential at the BZ edge, and the tapering scheme is depicted in Fig.~\ref{fig:4}(b). The variations across the tapering section comply with conventional minimum shot pitches~\cite{albrechtsen2022nanometer} in direct-write electron-beam lithography systems while providing sufficient adiabaticity to achieve ultra-low-loss cavities. To enhance the $Q$-factor, additional mirror sections with fixed $R = \SI{32}{nm}$ are added at both cavity edges. Typically, the $Q$-factor of such cavities rapidly increases with the number of mirrors, $M$, before eventually saturating. In our case, the $Q$-factor of the fundamental mode reaches its maximum value at $M=15$. The electric field amplitude distribution of this mode, occurring at $\lambda = \SI{1550}{nm}$, is shown in Fig.~\ref{fig:4}(c). The field exhibits hot spots in the waveguide defect region, the intensity of which drops from the center toward the edges of the tapered mirrors. Figure~\ref{fig:4}(d) illustrates the energy-density distribution of the fundamental mode along the axis of the waveguide and a Gaussian envelope, $f(x)$, fitted to the peak values. Given the slow variation of the envelope function at the scale of the lattice period and its Bloch-mode nature~\cite{charbonneau2002photonic}, the cavity mode volume can be approximated as
\begin{equation}\label{eq:V_mod_V_uc}
{V_{\text{mod}}}(\mathbf{r}_{\text{d}}) = \frac{\int_{\text{L}}dx |f(x)|^2}{a|f(x_{\text{d}})|^2}V_{\text{mod}}^{\text{UC}}(\mathbf{r}_{\text{d}}),
 \end{equation}
establishing a direct connection between the mode volume of the waveguide and the cavity~\cite{arregui2023cavity}. Employing Eq.~(\ref{eq:V_mod_V_uc}) alongside a direct numerical evaluation of Eq.~(\ref{eq:V_mod}) at the hot spot of the central bowtie, we achieve $V_{\text{mod}} = 6\times10^{-3}\lambda_{\text{res}}^3$, in good agreement with the value obtained by direct evaluation of the cavity version of Eq.~(\ref{eq:V_mod}) using the full cavity field, i.e.,  $V_{\text{mod}} = 5.9\times10^{-3}\lambda_{\text{res}}^3$. Formally, $V_{\text{mod}}$ in Eq.~(\ref{eq:V_mod}) has an additional surface term in case of the cavity due to its non-Hermitian character, but we disregard this term as its impact is negligible for cavities with high $Q$ factors~\cite{kristensen2020modeling}. We evaluate the Purcell factor for a resonant emitter polarized along the fundamental cavity mode and at its center,
\begin{equation}\label{eq:P_cav}
P(\mathbf{r}_{\text{d}}) = \frac{3}{4\pi^2}\left( \frac{\lambda}{n(\mathbf{r}_{\text{d}})}\right)^3\frac{Q}{V_{\text{mod}}(\mathbf{r}_{\text{d}})}
 \end{equation}
which results in $P = 3 \times 10^6$ owing to the large $Q$-factor and small $V_{\text{mod}}$. Compared to previously studied QVH cavity designs, which are ring cavities and exhibit a broad spatial field distribution extending over a significant area, our heterostructure design enables the confinement of the field to a much smaller region. This confinement leads to significantly reduced mode volume and, at the same time, a greatly increased $Q$-factor, which is three orders of magnitude higher than those calculated for ring cavities~\cite{smirnova2020room, noh2020experimental, zeng2020electrically, mehrabad2020chiral, gu2021topological, gong2020topological, liu2022topological}.

\section{\label{Conclusion} Conclusion}
In conclusion, the QVH PTI bowtie waveguide design presented here exhibits promising characteristics for enhancing light-matter interaction, achieving $P_x=70$ at $n_\text{g} = 30$ in the mid-plane of the slab. Compared to previous approaches~\cite{RevModPhys.87.347} to waveguide quantum electrodynamics, the attained high Purcell factor is not only achieved by a slow-light effect, i.e., a high $n_{\text{g}}$. Our work, therefore, shows a path to enhancing light-matter interaction beyond what is possible in conventional waveguides. Notably, both the slow-light effect and the bowtie effect are inherently broadband. However, the field concentration at the bowtie surfaces may also induce additional scattering, which calls for experimental investigations of propagation losses~\cite{rosiek_observation_2022, arora2023multiple}. In addition, the proposed waveguide design is multimodal, which may induce intramodal scattering, but the interplay between disorder and Anderson localization in bowtie waveguides has not yet been explored and would be an exciting avenue for future research. Further investigations could improve the dispersion by applying the topology-optimization techniques~\cite{christiansen2019designing}. Yet another important line of research should consider time-reversal-symmetry-broken systems, including magneto-optic materials~\cite{ni2023topological}, to achieve edge-mode non-reciprocity and, thus, ensure topological protection. 

We acknowledge financial support from the Villum Foundation Young Investigator Program (Grant No.\ 13170), Innovation Fund Denmark (Grant No.\ 0175-00022 - NEXUS and Grant No.\ 2054-00008 - SCALE), Independent Research Fund Denmark (Grant No.\ 0135-00315 - VAFL), the European Research Council (Grant No. 101045396 - SPOTLIGHT), and the Horizon Europe Research and Innovation Programme (Grant No.\ 101098961 - NEUROPIC and Grant no.\ 101067606 - TOPEX).

\bibliography{main}

\begin{thebibliography}{51}%
\makeatletter
\providecommand \@ifxundefined [1]{%
 \@ifx{#1\undefined}
}%
\providecommand \@ifnum [1]{%
 \ifnum #1\expandafter \@firstoftwo
 \else \expandafter \@secondoftwo
 \fi
}%
\providecommand \@ifx [1]{%
 \ifx #1\expandafter \@firstoftwo
 \else \expandafter \@secondoftwo
 \fi
}%
\providecommand \natexlab [1]{#1}%
\providecommand \enquote  [1]{``#1''}%
\providecommand \bibnamefont  [1]{#1}%
\providecommand \bibfnamefont [1]{#1}%
\providecommand \citenamefont [1]{#1}%
\providecommand \href@noop [0]{\@secondoftwo}%
\providecommand \href [0]{\begingroup \@sanitize@url \@href}%
\providecommand \@href[1]{\@@startlink{#1}\@@href}%
\providecommand \@@href[1]{\endgroup#1\@@endlink}%
\providecommand \@sanitize@url [0]{\catcode `\\12\catcode `\$12\catcode
  `\&12\catcode `\#12\catcode `\^12\catcode `\_12\catcode `\%12\relax}%
\providecommand \@@startlink[1]{}%
\providecommand \@@endlink[0]{}%
\providecommand \url  [0]{\begingroup\@sanitize@url \@url }%
\providecommand \@url [1]{\endgroup\@href {#1}{\urlprefix }}%
\providecommand \urlprefix  [0]{URL }%
\providecommand \Eprint [0]{\href }%
\providecommand \doibase [0]{https://doi.org/}%
\providecommand \selectlanguage [0]{\@gobble}%
\providecommand \bibinfo  [0]{\@secondoftwo}%
\providecommand \bibfield  [0]{\@secondoftwo}%
\providecommand \translation [1]{[#1]}%
\providecommand \BibitemOpen [0]{}%
\providecommand \bibitemStop [0]{}%
\providecommand \bibitemNoStop [0]{.\EOS\space}%
\providecommand \EOS [0]{\spacefactor3000\relax}%
\providecommand \BibitemShut  [1]{\csname bibitem#1\endcsname}%
\let\auto@bib@innerbib\@empty
\bibitem [{\citenamefont {Zhen}\ \emph {et~al.}(2013)\citenamefont {Zhen},
  \citenamefont {Chua}, \citenamefont {Lee}, \citenamefont {Rodriguez},
  \citenamefont {Liang}, \citenamefont {Johnson}, \citenamefont {Joannopoulos},
  \citenamefont {Solja{\v{c}}i{\'c}},\ and\ \citenamefont
  {Shapira}}]{zhen2013enabling}%
  \BibitemOpen
  \bibfield  {author} {\bibinfo {author} {\bibfnamefont {B.}~\bibnamefont
  {Zhen}}, \bibinfo {author} {\bibfnamefont {S.-L.}\ \bibnamefont {Chua}},
  \bibinfo {author} {\bibfnamefont {J.}~\bibnamefont {Lee}}, \bibinfo {author}
  {\bibfnamefont {A.~W.}\ \bibnamefont {Rodriguez}}, \bibinfo {author}
  {\bibfnamefont {X.}~\bibnamefont {Liang}}, \bibinfo {author} {\bibfnamefont
  {S.~G.}\ \bibnamefont {Johnson}}, \bibinfo {author} {\bibfnamefont {J.~D.}\
  \bibnamefont {Joannopoulos}}, \bibinfo {author} {\bibfnamefont
  {M.}~\bibnamefont {Solja{\v{c}}i{\'c}}},\ and\ \bibinfo {author}
  {\bibfnamefont {O.}~\bibnamefont {Shapira}},\ }\bibfield  {title} {\bibinfo
  {title} {Enabling enhanced emission and low-threshold lasing of organic
  molecules using special fano resonances of macroscopic photonic crystals},\
  }\href@noop {} {\bibfield  {journal} {\bibinfo  {journal} {PNAS}\ }\textbf
  {\bibinfo {volume} {110}},\ \bibinfo {pages} {13711} (\bibinfo {year}
  {2013})}\BibitemShut {NoStop}%
\bibitem [{\citenamefont {Altug}\ \emph {et~al.}(2022)\citenamefont {Altug},
  \citenamefont {Oh}, \citenamefont {Maier},\ and\ \citenamefont
  {Homola}}]{altug2022advances}%
  \BibitemOpen
  \bibfield  {author} {\bibinfo {author} {\bibfnamefont {H.}~\bibnamefont
  {Altug}}, \bibinfo {author} {\bibfnamefont {S.-H.}\ \bibnamefont {Oh}},
  \bibinfo {author} {\bibfnamefont {S.~A.}\ \bibnamefont {Maier}},\ and\
  \bibinfo {author} {\bibfnamefont {J.}~\bibnamefont {Homola}},\ }\bibfield
  {title} {\bibinfo {title} {Advances and applications of nanophotonic
  biosensors},\ }\href@noop {} {\bibfield  {journal} {\bibinfo  {journal} {Nat.
  Nanotechnol.}\ }\textbf {\bibinfo {volume} {17}},\ \bibinfo {pages} {5}
  (\bibinfo {year} {2022})}\BibitemShut {NoStop}%
\bibitem [{\citenamefont {Beliaev}\ \emph {et~al.}(2022)\citenamefont
  {Beliaev}, \citenamefont {Stounbjerg}, \citenamefont {Finco}, \citenamefont
  {Bunea}, \citenamefont {Malureanu}, \citenamefont {Lindvold}, \citenamefont
  {Takayama}, \citenamefont {Andersen},\ and\ \citenamefont
  {Lavrinenko}}]{beliaev2022pedestal}%
  \BibitemOpen
  \bibfield  {author} {\bibinfo {author} {\bibfnamefont {L.~Y.}\ \bibnamefont
  {Beliaev}}, \bibinfo {author} {\bibfnamefont {P.~G.}\ \bibnamefont
  {Stounbjerg}}, \bibinfo {author} {\bibfnamefont {G.}~\bibnamefont {Finco}},
  \bibinfo {author} {\bibfnamefont {A.-I.}\ \bibnamefont {Bunea}}, \bibinfo
  {author} {\bibfnamefont {R.}~\bibnamefont {Malureanu}}, \bibinfo {author}
  {\bibfnamefont {L.~R.}\ \bibnamefont {Lindvold}}, \bibinfo {author}
  {\bibfnamefont {O.}~\bibnamefont {Takayama}}, \bibinfo {author}
  {\bibfnamefont {P.~E.}\ \bibnamefont {Andersen}},\ and\ \bibinfo {author}
  {\bibfnamefont {A.~V.}\ \bibnamefont {Lavrinenko}},\ }\bibfield  {title}
  {\bibinfo {title} {Pedestal high-contrast gratings for biosensing},\
  }\href@noop {} {\bibfield  {journal} {\bibinfo  {journal} {Nanomater.}\
  }\textbf {\bibinfo {volume} {12}},\ \bibinfo {pages} {1748} (\bibinfo {year}
  {2022})}\BibitemShut {NoStop}%
\bibitem [{\citenamefont {Lodahl}\ \emph {et~al.}(2015)\citenamefont {Lodahl},
  \citenamefont {Mahmoodian},\ and\ \citenamefont
  {Stobbe}}]{RevModPhys.87.347}%
  \BibitemOpen
  \bibfield  {author} {\bibinfo {author} {\bibfnamefont {P.}~\bibnamefont
  {Lodahl}}, \bibinfo {author} {\bibfnamefont {S.}~\bibnamefont {Mahmoodian}},\
  and\ \bibinfo {author} {\bibfnamefont {S.}~\bibnamefont {Stobbe}},\
  }\bibfield  {title} {\bibinfo {title} {Interfacing single photons and single
  quantum dots with photonic nanostructures},\ }\href
  {https://doi.org/10.1103/RevModPhys.87.347} {\bibfield  {journal} {\bibinfo
  {journal} {Rev. Mod. Phys.}\ }\textbf {\bibinfo {volume} {87}},\ \bibinfo
  {pages} {347} (\bibinfo {year} {2015})}\BibitemShut {NoStop}%
\bibitem [{\citenamefont {Romeira}\ and\ \citenamefont
  {Fiore}(2018)}]{romeira2018purcell}%
  \BibitemOpen
  \bibfield  {author} {\bibinfo {author} {\bibfnamefont {B.}~\bibnamefont
  {Romeira}}\ and\ \bibinfo {author} {\bibfnamefont {A.}~\bibnamefont
  {Fiore}},\ }\bibfield  {title} {\bibinfo {title} {Purcell effect in the
  stimulated and spontaneous emission rates of nanoscale semiconductor
  lasers},\ }\href@noop {} {\bibfield  {journal} {\bibinfo  {journal} {IEEE J.
  Quantum Electron.}\ }\textbf {\bibinfo {volume} {54}},\ \bibinfo {pages} {1}
  (\bibinfo {year} {2018})}\BibitemShut {NoStop}%
\bibitem [{\citenamefont {Lopez}(2003)}]{lopez2003materials}%
  \BibitemOpen
  \bibfield  {author} {\bibinfo {author} {\bibfnamefont {C.}~\bibnamefont
  {Lopez}},\ }\bibfield  {title} {\bibinfo {title} {Materials aspects of
  photonic crystals},\ }\href@noop {} {\bibfield  {journal} {\bibinfo
  {journal} {Adv. Mater.}\ }\textbf {\bibinfo {volume} {15}},\ \bibinfo {pages}
  {1679} (\bibinfo {year} {2003})}\BibitemShut {NoStop}%
\bibitem [{\citenamefont {Ozawa}\ \emph {et~al.}(2019)\citenamefont {Ozawa},
  \citenamefont {Price}, \citenamefont {Amo}, \citenamefont {Goldman},
  \citenamefont {Hafezi}, \citenamefont {Lu}, \citenamefont {Rechtsman},
  \citenamefont {Schuster}, \citenamefont {Simon}, \citenamefont {Zilberberg},\
  and\ \citenamefont {Carusotto}}]{ozawa_topological_2019}%
  \BibitemOpen
  \bibfield  {author} {\bibinfo {author} {\bibfnamefont {T.}~\bibnamefont
  {Ozawa}}, \bibinfo {author} {\bibfnamefont {H.~M.}\ \bibnamefont {Price}},
  \bibinfo {author} {\bibfnamefont {A.}~\bibnamefont {Amo}}, \bibinfo {author}
  {\bibfnamefont {N.}~\bibnamefont {Goldman}}, \bibinfo {author} {\bibfnamefont
  {M.}~\bibnamefont {Hafezi}}, \bibinfo {author} {\bibfnamefont
  {L.}~\bibnamefont {Lu}}, \bibinfo {author} {\bibfnamefont {M.~C.}\
  \bibnamefont {Rechtsman}}, \bibinfo {author} {\bibfnamefont {D.}~\bibnamefont
  {Schuster}}, \bibinfo {author} {\bibfnamefont {J.}~\bibnamefont {Simon}},
  \bibinfo {author} {\bibfnamefont {O.}~\bibnamefont {Zilberberg}},\ and\
  \bibinfo {author} {\bibfnamefont {I.}~\bibnamefont {Carusotto}},\ }\bibfield
  {title} {\bibinfo {title} {Topological photonics},\ }\href
  {https://doi.org/10.1103/RevModPhys.91.015006} {\bibfield  {journal}
  {\bibinfo  {journal} {Rev. Mod. Phys.}\ }\textbf {\bibinfo {volume} {91}},\
  \bibinfo {pages} {015006} (\bibinfo {year} {2019})}\BibitemShut {NoStop}%
\bibitem [{\citenamefont {Kim}\ \emph {et~al.}(2020)\citenamefont {Kim},
  \citenamefont {Jacob},\ and\ \citenamefont {Rho}}]{kim2020recent}%
  \BibitemOpen
  \bibfield  {author} {\bibinfo {author} {\bibfnamefont {M.}~\bibnamefont
  {Kim}}, \bibinfo {author} {\bibfnamefont {Z.}~\bibnamefont {Jacob}},\ and\
  \bibinfo {author} {\bibfnamefont {J.}~\bibnamefont {Rho}},\ }\bibfield
  {title} {\bibinfo {title} {Recent advances in 2d, 3d and higher-order
  topological photonics},\ }\href@noop {} {\bibfield  {journal} {\bibinfo
  {journal} {Light Sci. Appl.}\ }\textbf {\bibinfo {volume} {9}},\ \bibinfo
  {pages} {130} (\bibinfo {year} {2020})}\BibitemShut {NoStop}%
\bibitem [{\citenamefont {Ni}\ \emph {et~al.}(2023)\citenamefont {Ni},
  \citenamefont {Yves}, \citenamefont {Krasnok},\ and\ \citenamefont
  {Alu}}]{ni2023topological}%
  \BibitemOpen
  \bibfield  {author} {\bibinfo {author} {\bibfnamefont {X.}~\bibnamefont
  {Ni}}, \bibinfo {author} {\bibfnamefont {S.}~\bibnamefont {Yves}}, \bibinfo
  {author} {\bibfnamefont {A.}~\bibnamefont {Krasnok}},\ and\ \bibinfo {author}
  {\bibfnamefont {A.}~\bibnamefont {Alu}},\ }\bibfield  {title} {\bibinfo
  {title} {Topological metamaterials},\ }\href@noop {} {\bibfield  {journal}
  {\bibinfo  {journal} {Chem. Rev.}\ }\textbf {\bibinfo {volume} {123}},\
  \bibinfo {pages} {7585} (\bibinfo {year} {2023})}\BibitemShut {NoStop}%
\bibitem [{\citenamefont {Ma}\ and\ \citenamefont {Shvets}(2016)}]{Ma_2016}%
  \BibitemOpen
  \bibfield  {author} {\bibinfo {author} {\bibfnamefont {T.}~\bibnamefont
  {Ma}}\ and\ \bibinfo {author} {\bibfnamefont {G.}~\bibnamefont {Shvets}},\
  }\bibfield  {title} {\bibinfo {title} {All-{Si} valley-{Hall} photonic
  topological insulator},\ }\href
  {https://doi.org/10.1088/1367-2630/18/2/025012} {\bibfield  {journal}
  {\bibinfo  {journal} {New J. Phys.}\ }\textbf {\bibinfo {volume} {18}},\
  \bibinfo {pages} {025012} (\bibinfo {year} {2016})}\BibitemShut {NoStop}%
\bibitem [{\citenamefont {Yoshimi}\ \emph {et~al.}(2020)\citenamefont
  {Yoshimi}, \citenamefont {Yamaguchi}, \citenamefont {Ota}, \citenamefont
  {Arakawa},\ and\ \citenamefont {Iwamoto}}]{yoshimi2020slow}%
  \BibitemOpen
  \bibfield  {author} {\bibinfo {author} {\bibfnamefont {H.}~\bibnamefont
  {Yoshimi}}, \bibinfo {author} {\bibfnamefont {T.}~\bibnamefont {Yamaguchi}},
  \bibinfo {author} {\bibfnamefont {Y.}~\bibnamefont {Ota}}, \bibinfo {author}
  {\bibfnamefont {Y.}~\bibnamefont {Arakawa}},\ and\ \bibinfo {author}
  {\bibfnamefont {S.}~\bibnamefont {Iwamoto}},\ }\bibfield  {title} {\bibinfo
  {title} {Slow light waveguides in topological valley photonic crystals},\
  }\href@noop {} {\bibfield  {journal} {\bibinfo  {journal} {Opt. Lett.}\
  }\textbf {\bibinfo {volume} {45}},\ \bibinfo {pages} {2648} (\bibinfo {year}
  {2020})}\BibitemShut {NoStop}%
\bibitem [{\citenamefont {Lodahl}\ \emph {et~al.}(2017)\citenamefont {Lodahl},
  \citenamefont {Mahmoodian}, \citenamefont {Stobbe}, \citenamefont
  {Rauschenbeutel}, \citenamefont {Schneeweiss}, \citenamefont {Volz},
  \citenamefont {Pichler},\ and\ \citenamefont {Zoller}}]{lodahl2017chiral}%
  \BibitemOpen
  \bibfield  {author} {\bibinfo {author} {\bibfnamefont {P.}~\bibnamefont
  {Lodahl}}, \bibinfo {author} {\bibfnamefont {S.}~\bibnamefont {Mahmoodian}},
  \bibinfo {author} {\bibfnamefont {S.}~\bibnamefont {Stobbe}}, \bibinfo
  {author} {\bibfnamefont {A.}~\bibnamefont {Rauschenbeutel}}, \bibinfo
  {author} {\bibfnamefont {P.}~\bibnamefont {Schneeweiss}}, \bibinfo {author}
  {\bibfnamefont {J.}~\bibnamefont {Volz}}, \bibinfo {author} {\bibfnamefont
  {H.}~\bibnamefont {Pichler}},\ and\ \bibinfo {author} {\bibfnamefont
  {P.}~\bibnamefont {Zoller}},\ }\bibfield  {title} {\bibinfo {title} {Chiral
  quantum optics},\ }\href@noop {} {\bibfield  {journal} {\bibinfo  {journal}
  {Nature}\ }\textbf {\bibinfo {volume} {541}},\ \bibinfo {pages} {473}
  (\bibinfo {year} {2017})}\BibitemShut {NoStop}%
\bibitem [{\citenamefont {Hauff}\ \emph {et~al.}(2022)\citenamefont {Hauff},
  \citenamefont {Le~Jeannic}, \citenamefont {Lodahl}, \citenamefont {Hughes},\
  and\ \citenamefont {Rotenberg}}]{hauff2022chiral}%
  \BibitemOpen
  \bibfield  {author} {\bibinfo {author} {\bibfnamefont {N.~V.}\ \bibnamefont
  {Hauff}}, \bibinfo {author} {\bibfnamefont {H.}~\bibnamefont {Le~Jeannic}},
  \bibinfo {author} {\bibfnamefont {P.}~\bibnamefont {Lodahl}}, \bibinfo
  {author} {\bibfnamefont {S.}~\bibnamefont {Hughes}},\ and\ \bibinfo {author}
  {\bibfnamefont {N.}~\bibnamefont {Rotenberg}},\ }\bibfield  {title} {\bibinfo
  {title} {Chiral quantum optics in broken-symmetry and topological photonic
  crystal waveguides},\ }\href@noop {} {\bibfield  {journal} {\bibinfo
  {journal} {Phys. Rev. Res.}\ }\textbf {\bibinfo {volume} {4}},\ \bibinfo
  {pages} {023082} (\bibinfo {year} {2022})}\BibitemShut {NoStop}%
\bibitem [{\citenamefont {Arregui}\ \emph {et~al.}(2021)\citenamefont
  {Arregui}, \citenamefont {Gomis-Bresco}, \citenamefont {Sotomayor-Torres},\
  and\ \citenamefont {Garcia}}]{Bravo_2021}%
  \BibitemOpen
  \bibfield  {author} {\bibinfo {author} {\bibfnamefont {G.}~\bibnamefont
  {Arregui}}, \bibinfo {author} {\bibfnamefont {J.}~\bibnamefont
  {Gomis-Bresco}}, \bibinfo {author} {\bibfnamefont {C.~M.}\ \bibnamefont
  {Sotomayor-Torres}},\ and\ \bibinfo {author} {\bibfnamefont {P.~D.}\
  \bibnamefont {Garcia}},\ }\bibfield  {title} {\bibinfo {title} {Quantifying
  the robustness of topological slow light},\ }\href@noop {} {\bibfield
  {journal} {\bibinfo  {journal} {Phys. Rev. Lett.}\ }\textbf {\bibinfo
  {volume} {126}},\ \bibinfo {pages} {027403} (\bibinfo {year}
  {2021})}\BibitemShut {NoStop}%
\bibitem [{\citenamefont {Rosiek}\ \emph {et~al.}(2023)\citenamefont {Rosiek},
  \citenamefont {Arregui}, \citenamefont {Vladimirova}, \citenamefont
  {Albrechtsen}, \citenamefont {Vongughi~Lahijani}, \citenamefont
  {Christiansen},\ and\ \citenamefont {Stobbe}}]{rosiek_observation_2022}%
  \BibitemOpen
  \bibfield  {author} {\bibinfo {author} {\bibfnamefont {C.~A.}\ \bibnamefont
  {Rosiek}}, \bibinfo {author} {\bibfnamefont {G.}~\bibnamefont {Arregui}},
  \bibinfo {author} {\bibfnamefont {A.}~\bibnamefont {Vladimirova}}, \bibinfo
  {author} {\bibfnamefont {M.}~\bibnamefont {Albrechtsen}}, \bibinfo {author}
  {\bibfnamefont {B.}~\bibnamefont {Vongughi~Lahijani}}, \bibinfo {author}
  {\bibfnamefont {R.~E.}\ \bibnamefont {Christiansen}},\ and\ \bibinfo {author}
  {\bibfnamefont {S.}~\bibnamefont {Stobbe}},\ }\bibfield  {title} {\bibinfo
  {title} {Observation of strong backscattering in valley-{Hall} topological
  interface modes},\ }\href {https://doi.org/10.1038/s41566-023-01189-x}
  {\bibfield  {journal} {\bibinfo  {journal} {Nat. Photon.}\ }\textbf {\bibinfo
  {volume} {17}},\ \bibinfo {pages} {386} (\bibinfo {year} {2023})}\BibitemShut
  {NoStop}%
\bibitem [{\citenamefont {Arora}\ \emph {et~al.}(2023)\citenamefont {Arora},
  \citenamefont {Bauer}, \citenamefont {Barczyk}, \citenamefont {Verhagen},\
  and\ \citenamefont {Kuipers}}]{arora2023multiple}%
  \BibitemOpen
  \bibfield  {author} {\bibinfo {author} {\bibfnamefont {S.}~\bibnamefont
  {Arora}}, \bibinfo {author} {\bibfnamefont {T.}~\bibnamefont {Bauer}},
  \bibinfo {author} {\bibfnamefont {R.}~\bibnamefont {Barczyk}}, \bibinfo
  {author} {\bibfnamefont {E.}~\bibnamefont {Verhagen}},\ and\ \bibinfo
  {author} {\bibfnamefont {L.}~\bibnamefont {Kuipers}},\ }\href@noop {}
  {\bibinfo {title} {Multiple backscattering in trivial and non-trivial
  topological photonic crystal edge states with controlled disorder}} (\bibinfo
  {year} {2023}),\ \Eprint {https://arxiv.org/abs/2310.02978v2} {2310.02978v2
  [physics.optics]} \BibitemShut {NoStop}%
\bibitem [{\citenamefont {Gondarenko}\ \emph {et~al.}(2006)\citenamefont
  {Gondarenko}, \citenamefont {Preble}, \citenamefont {Robinson}, \citenamefont
  {Chen}, \citenamefont {Lipson},\ and\ \citenamefont
  {Lipson}}]{gondarenko2006spontaneous}%
  \BibitemOpen
  \bibfield  {author} {\bibinfo {author} {\bibfnamefont {A.}~\bibnamefont
  {Gondarenko}}, \bibinfo {author} {\bibfnamefont {S.}~\bibnamefont {Preble}},
  \bibinfo {author} {\bibfnamefont {J.}~\bibnamefont {Robinson}}, \bibinfo
  {author} {\bibfnamefont {L.}~\bibnamefont {Chen}}, \bibinfo {author}
  {\bibfnamefont {H.}~\bibnamefont {Lipson}},\ and\ \bibinfo {author}
  {\bibfnamefont {M.}~\bibnamefont {Lipson}},\ }\bibfield  {title} {\bibinfo
  {title} {Spontaneous emergence of periodic patterns in a biologically
  inspired simulation of photonic structures},\ }\href@noop {} {\bibfield
  {journal} {\bibinfo  {journal} {Phys. Rev. Lett.}\ }\textbf {\bibinfo
  {volume} {96}},\ \bibinfo {pages} {143904} (\bibinfo {year}
  {2006})}\BibitemShut {NoStop}%
\bibitem [{\citenamefont {Hu}\ and\ \citenamefont
  {Weiss}(2016)}]{hu2016design}%
  \BibitemOpen
  \bibfield  {author} {\bibinfo {author} {\bibfnamefont {S.}~\bibnamefont
  {Hu}}\ and\ \bibinfo {author} {\bibfnamefont {S.~M.}\ \bibnamefont {Weiss}},\
  }\bibfield  {title} {\bibinfo {title} {Design of photonic crystal cavities
  for extreme light concentration},\ }\href@noop {} {\bibfield  {journal}
  {\bibinfo  {journal} {ACS Photon.}\ }\textbf {\bibinfo {volume} {3}},\
  \bibinfo {pages} {1647} (\bibinfo {year} {2016})}\BibitemShut {NoStop}%
\bibitem [{\citenamefont {Choi}\ \emph {et~al.}(2017)\citenamefont {Choi},
  \citenamefont {Heuck},\ and\ \citenamefont {Englund}}]{choi2017self}%
  \BibitemOpen
  \bibfield  {author} {\bibinfo {author} {\bibfnamefont {H.}~\bibnamefont
  {Choi}}, \bibinfo {author} {\bibfnamefont {M.}~\bibnamefont {Heuck}},\ and\
  \bibinfo {author} {\bibfnamefont {D.}~\bibnamefont {Englund}},\ }\bibfield
  {title} {\bibinfo {title} {Self-similar nanocavity design with ultrasmall
  mode volume for single-photon nonlinearities},\ }\href@noop {} {\bibfield
  {journal} {\bibinfo  {journal} {Phys. Rev. Lett.}\ }\textbf {\bibinfo
  {volume} {118}},\ \bibinfo {pages} {223605} (\bibinfo {year}
  {2017})}\BibitemShut {NoStop}%
\bibitem [{\citenamefont {Wang}\ \emph {et~al.}(2018)\citenamefont {Wang},
  \citenamefont {Christiansen}, \citenamefont {Yu}, \citenamefont {M{\o}rk},\
  and\ \citenamefont {Sigmund}}]{wang2018maximizing}%
  \BibitemOpen
  \bibfield  {author} {\bibinfo {author} {\bibfnamefont {F.}~\bibnamefont
  {Wang}}, \bibinfo {author} {\bibfnamefont {R.~E.}\ \bibnamefont
  {Christiansen}}, \bibinfo {author} {\bibfnamefont {Y.}~\bibnamefont {Yu}},
  \bibinfo {author} {\bibfnamefont {J.}~\bibnamefont {M{\o}rk}},\ and\ \bibinfo
  {author} {\bibfnamefont {O.}~\bibnamefont {Sigmund}},\ }\bibfield  {title}
  {\bibinfo {title} {Maximizing the quality factor to mode volume ratio for
  ultra-small photonic crystal cavities},\ }\href@noop {} {\bibfield  {journal}
  {\bibinfo  {journal} {Appl. Phys. Lett.}\ }\textbf {\bibinfo {volume} {113}}
  (\bibinfo {year} {2018})}\BibitemShut {NoStop}%
\bibitem [{\citenamefont {Christiansen}\ \emph {et~al.}(2023)\citenamefont
  {Christiansen}, \citenamefont {Kristensen}, \citenamefont {M{\o}rk},\ and\
  \citenamefont {Sigmund}}]{christiansen2023impact}%
  \BibitemOpen
  \bibfield  {author} {\bibinfo {author} {\bibfnamefont {R.~E.}\ \bibnamefont
  {Christiansen}}, \bibinfo {author} {\bibfnamefont {P.~T.}\ \bibnamefont
  {Kristensen}}, \bibinfo {author} {\bibfnamefont {J.}~\bibnamefont
  {M{\o}rk}},\ and\ \bibinfo {author} {\bibfnamefont {O.}~\bibnamefont
  {Sigmund}},\ }\bibfield  {title} {\bibinfo {title} {Impact of figures of
  merit in photonic inverse design},\ }\href@noop {} {\bibfield  {journal}
  {\bibinfo  {journal} {Opt. Express}\ }\textbf {\bibinfo {volume} {31}},\
  \bibinfo {pages} {8363} (\bibinfo {year} {2023})}\BibitemShut {NoStop}%
\bibitem [{\citenamefont {Albrechtsen}\ \emph
  {et~al.}(2022{\natexlab{a}})\citenamefont {Albrechtsen}, \citenamefont
  {Vosoughi~Lahijani}, \citenamefont {Christiansen}, \citenamefont {Nguyen},
  \citenamefont {Casses}, \citenamefont {Hansen}, \citenamefont {Stenger},
  \citenamefont {Sigmund}, \citenamefont {Jansen}, \citenamefont {M{\o}rk},\
  and\ \citenamefont {Stobbe}}]{albrechtsen2022nanometer}%
  \BibitemOpen
  \bibfield  {author} {\bibinfo {author} {\bibfnamefont {M.}~\bibnamefont
  {Albrechtsen}}, \bibinfo {author} {\bibfnamefont {B.}~\bibnamefont
  {Vosoughi~Lahijani}}, \bibinfo {author} {\bibfnamefont {R.~E.}\ \bibnamefont
  {Christiansen}}, \bibinfo {author} {\bibfnamefont {V.~T.~H.}\ \bibnamefont
  {Nguyen}}, \bibinfo {author} {\bibfnamefont {L.~N.}\ \bibnamefont {Casses}},
  \bibinfo {author} {\bibfnamefont {S.~E.}\ \bibnamefont {Hansen}}, \bibinfo
  {author} {\bibfnamefont {N.}~\bibnamefont {Stenger}}, \bibinfo {author}
  {\bibfnamefont {O.}~\bibnamefont {Sigmund}}, \bibinfo {author} {\bibfnamefont
  {H.}~\bibnamefont {Jansen}}, \bibinfo {author} {\bibfnamefont
  {J.}~\bibnamefont {M{\o}rk}},\ and\ \bibinfo {author} {\bibfnamefont
  {S.}~\bibnamefont {Stobbe}},\ }\bibfield  {title} {\bibinfo {title}
  {Nanometer-scale photon confinement in topology-optimized dielectric
  cavities},\ }\href@noop {} {\bibfield  {journal} {\bibinfo  {journal} {Nat.
  Commun.}\ }\textbf {\bibinfo {volume} {13}},\ \bibinfo {pages} {6281}
  (\bibinfo {year} {2022}{\natexlab{a}})}\BibitemShut {NoStop}%
\bibitem [{\citenamefont {Babar}\ \emph {et~al.}(2023)\citenamefont {Babar},
  \citenamefont {Weis}, \citenamefont {Tsoukalas}, \citenamefont
  {Kadkhodazadeh}, \citenamefont {Arregui}, \citenamefont {Vosoughi~Lahijani},\
  and\ \citenamefont {Stobbe}}]{babar2023self}%
  \BibitemOpen
  \bibfield  {author} {\bibinfo {author} {\bibfnamefont {A.~N.}\ \bibnamefont
  {Babar}}, \bibinfo {author} {\bibfnamefont {T.}~\bibnamefont {Weis}},
  \bibinfo {author} {\bibfnamefont {K.}~\bibnamefont {Tsoukalas}}, \bibinfo
  {author} {\bibfnamefont {S.}~\bibnamefont {Kadkhodazadeh}}, \bibinfo {author}
  {\bibfnamefont {G.}~\bibnamefont {Arregui}}, \bibinfo {author} {\bibfnamefont
  {B.}~\bibnamefont {Vosoughi~Lahijani}},\ and\ \bibinfo {author}
  {\bibfnamefont {S.}~\bibnamefont {Stobbe}},\ }\bibfield  {title} {\bibinfo
  {title} {Self-assembled photonic cavities with atomic-scale confinement},\
  }\href@noop {} {\bibfield  {journal} {\bibinfo  {journal} {Nature}\ }\textbf
  {\bibinfo {volume} {624}},\ \bibinfo {pages} {57} (\bibinfo {year}
  {2023})}\BibitemShut {NoStop}%
\bibitem [{\citenamefont {Purcell}(1995)}]{purcell1995spontaneous}%
  \BibitemOpen
  \bibfield  {author} {\bibinfo {author} {\bibfnamefont {E.~M.}\ \bibnamefont
  {Purcell}},\ }\bibfield  {title} {\bibinfo {title} {Spontaneous emission
  probabilities at radio frequencies},\ }in\ \href@noop {} {\emph {\bibinfo
  {booktitle} {Confined Electrons and Photons: New Physics and Applications}}}\
  (\bibinfo {year} {1995})\ pp.\ \bibinfo {pages} {839--839}\BibitemShut
  {NoStop}%
\bibitem [{\citenamefont {Xue}\ \emph {et~al.}()\citenamefont {Xue},
  \citenamefont {Yang},\ and\ \citenamefont {Zhang}}]{VH_review_2018}%
  \BibitemOpen
  \bibfield  {author} {\bibinfo {author} {\bibfnamefont {H.}~\bibnamefont
  {Xue}}, \bibinfo {author} {\bibfnamefont {Y.}~\bibnamefont {Yang}},\ and\
  \bibinfo {author} {\bibfnamefont {B.}~\bibnamefont {Zhang}},\ }\bibfield
  {title} {\bibinfo {title} {Topological valley photonics: Physics and device
  applications},\ }\href
  {https://doi.org/https://doi.org/10.1002/adpr.202100013} {\bibfield
  {journal} {\bibinfo  {journal} {Adv. Photonics Res.}\ }\textbf {\bibinfo
  {volume} {2}},\ \bibinfo {pages} {2100013}}\BibitemShut {NoStop}%
\bibitem [{\citenamefont {Albrechtsen}\ \emph
  {et~al.}(2022{\natexlab{b}})\citenamefont {Albrechtsen}, \citenamefont
  {Vosoughi~Lahijani},\ and\ \citenamefont {Stobbe}}]{albrechtsen2022two}%
  \BibitemOpen
  \bibfield  {author} {\bibinfo {author} {\bibfnamefont {M.}~\bibnamefont
  {Albrechtsen}}, \bibinfo {author} {\bibfnamefont {B.}~\bibnamefont
  {Vosoughi~Lahijani}},\ and\ \bibinfo {author} {\bibfnamefont
  {S.}~\bibnamefont {Stobbe}},\ }\bibfield  {title} {\bibinfo {title} {Two
  regimes of confinement in photonic nanocavities: bulk confinement versus
  lightning rods},\ }\href@noop {} {\bibfield  {journal} {\bibinfo  {journal}
  {Opt. Express}\ }\textbf {\bibinfo {volume} {30}},\ \bibinfo {pages} {15458}
  (\bibinfo {year} {2022}{\natexlab{b}})}\BibitemShut {NoStop}%
\bibitem [{\citenamefont {Shalaev}\ \emph {et~al.}(2019)\citenamefont
  {Shalaev}, \citenamefont {Walasik}, \citenamefont {Tsukernik}, \citenamefont
  {Xu},\ and\ \citenamefont {Litchinitser}}]{shalaev2019robust}%
  \BibitemOpen
  \bibfield  {author} {\bibinfo {author} {\bibfnamefont {M.~I.}\ \bibnamefont
  {Shalaev}}, \bibinfo {author} {\bibfnamefont {W.}~\bibnamefont {Walasik}},
  \bibinfo {author} {\bibfnamefont {A.}~\bibnamefont {Tsukernik}}, \bibinfo
  {author} {\bibfnamefont {Y.}~\bibnamefont {Xu}},\ and\ \bibinfo {author}
  {\bibfnamefont {N.~M.}\ \bibnamefont {Litchinitser}},\ }\bibfield  {title}
  {\bibinfo {title} {Robust topologically protected transport in photonic
  crystals at telecommunication wavelengths},\ }\href@noop {} {\bibfield
  {journal} {\bibinfo  {journal} {Nat. Nanotechnol.}\ }\textbf {\bibinfo
  {volume} {14}},\ \bibinfo {pages} {31} (\bibinfo {year} {2019})}\BibitemShut
  {NoStop}%
\bibitem [{\citenamefont {Hughes}(2004)}]{Purcell-Hughes-2004}%
  \BibitemOpen
  \bibfield  {author} {\bibinfo {author} {\bibfnamefont {S.}~\bibnamefont
  {Hughes}},\ }\bibfield  {title} {\bibinfo {title} {Enhanced single-photon
  emission from quantum dots in photonic crystal waveguides and nanocavities},\
  }\href {https://doi.org/10.1364/OL.29.002659} {\bibfield  {journal} {\bibinfo
   {journal} {Opt. Lett.}\ }\textbf {\bibinfo {volume} {29}},\ \bibinfo {pages}
  {2659} (\bibinfo {year} {2004})}\BibitemShut {NoStop}%
\bibitem [{\citenamefont {Rao}\ and\ \citenamefont
  {Hughes}(2007)}]{rao2007single}%
  \BibitemOpen
  \bibfield  {author} {\bibinfo {author} {\bibfnamefont {V.~M.}\ \bibnamefont
  {Rao}}\ and\ \bibinfo {author} {\bibfnamefont {S.}~\bibnamefont {Hughes}},\
  }\bibfield  {title} {\bibinfo {title} {Single quantum-dot purcell factor and
  $\beta$ factor in a photonic crystal waveguide},\ }\href@noop {} {\bibfield
  {journal} {\bibinfo  {journal} {Phys. Rev. B}\ }\textbf {\bibinfo {volume}
  {75}},\ \bibinfo {pages} {205437} (\bibinfo {year} {2007})}\BibitemShut
  {NoStop}%
\bibitem [{\citenamefont {Arregui}\ \emph {et~al.}(2023)\citenamefont
  {Arregui}, \citenamefont {Ng}, \citenamefont {Albrechtsen}, \citenamefont
  {Stobbe}, \citenamefont {Sotomayor-Torres},\ and\ \citenamefont
  {Garc{\'\i}a}}]{arregui2023cavity}%
  \BibitemOpen
  \bibfield  {author} {\bibinfo {author} {\bibfnamefont {G.}~\bibnamefont
  {Arregui}}, \bibinfo {author} {\bibfnamefont {R.~C.}\ \bibnamefont {Ng}},
  \bibinfo {author} {\bibfnamefont {M.}~\bibnamefont {Albrechtsen}}, \bibinfo
  {author} {\bibfnamefont {S.}~\bibnamefont {Stobbe}}, \bibinfo {author}
  {\bibfnamefont {C.}~\bibnamefont {Sotomayor-Torres}},\ and\ \bibinfo {author}
  {\bibfnamefont {P.~D.}\ \bibnamefont {Garc{\'\i}a}},\ }\bibfield  {title}
  {\bibinfo {title} {Cavity optomechanics with anderson-localized optical
  modes},\ }\href@noop {} {\bibfield  {journal} {\bibinfo  {journal} {Phys.
  Rev. Lett.}\ }\textbf {\bibinfo {volume} {130}},\ \bibinfo {pages} {043802}
  (\bibinfo {year} {2023})}\BibitemShut {NoStop}%
\bibitem [{\citenamefont {Garc{\'\i}a}\ and\ \citenamefont
  {Lodahl}(2017)}]{garcia2017physics}%
  \BibitemOpen
  \bibfield  {author} {\bibinfo {author} {\bibfnamefont {P.~D.}\ \bibnamefont
  {Garc{\'\i}a}}\ and\ \bibinfo {author} {\bibfnamefont {P.}~\bibnamefont
  {Lodahl}},\ }\bibfield  {title} {\bibinfo {title} {Physics of quantum light
  emitters in disordered photonic nanostructures},\ }\href@noop {} {\bibfield
  {journal} {\bibinfo  {journal} {Ann. Phys.}\ }\textbf {\bibinfo {volume}
  {529}},\ \bibinfo {pages} {1600351} (\bibinfo {year} {2017})}\BibitemShut
  {NoStop}%
\bibitem [{\citenamefont {Polman}(1997)}]{polman1997erbium}%
  \BibitemOpen
  \bibfield  {author} {\bibinfo {author} {\bibfnamefont {A.}~\bibnamefont
  {Polman}},\ }\bibfield  {title} {\bibinfo {title} {Erbium implanted thin film
  photonic materials},\ }\href@noop {} {\bibfield  {journal} {\bibinfo
  {journal} {J. Appl. Phys.}\ }\textbf {\bibinfo {volume} {82}},\ \bibinfo
  {pages} {1} (\bibinfo {year} {1997})}\BibitemShut {NoStop}%
\bibitem [{\citenamefont {Weiss}\ \emph {et~al.}(2021)\citenamefont {Weiss},
  \citenamefont {Gritsch}, \citenamefont {Merkel},\ and\ \citenamefont
  {Reiserer}}]{weiss2021erbium}%
  \BibitemOpen
  \bibfield  {author} {\bibinfo {author} {\bibfnamefont {L.}~\bibnamefont
  {Weiss}}, \bibinfo {author} {\bibfnamefont {A.}~\bibnamefont {Gritsch}},
  \bibinfo {author} {\bibfnamefont {B.}~\bibnamefont {Merkel}},\ and\ \bibinfo
  {author} {\bibfnamefont {A.}~\bibnamefont {Reiserer}},\ }\bibfield  {title}
  {\bibinfo {title} {Erbium dopants in nanophotonic silicon waveguides},\
  }\href@noop {} {\bibfield  {journal} {\bibinfo  {journal} {Optica}\ }\textbf
  {\bibinfo {volume} {8}},\ \bibinfo {pages} {40} (\bibinfo {year}
  {2021})}\BibitemShut {NoStop}%
\bibitem [{\citenamefont {Gritsch}\ \emph {et~al.}(2022)\citenamefont
  {Gritsch}, \citenamefont {Weiss}, \citenamefont {Fr{\"u}h}, \citenamefont
  {Rinner},\ and\ \citenamefont {Reiserer}}]{gritsch2022narrow}%
  \BibitemOpen
  \bibfield  {author} {\bibinfo {author} {\bibfnamefont {A.}~\bibnamefont
  {Gritsch}}, \bibinfo {author} {\bibfnamefont {L.}~\bibnamefont {Weiss}},
  \bibinfo {author} {\bibfnamefont {J.}~\bibnamefont {Fr{\"u}h}}, \bibinfo
  {author} {\bibfnamefont {S.}~\bibnamefont {Rinner}},\ and\ \bibinfo {author}
  {\bibfnamefont {A.}~\bibnamefont {Reiserer}},\ }\bibfield  {title} {\bibinfo
  {title} {Narrow optical transitions in erbium-implanted silicon waveguides},\
  }\href@noop {} {\bibfield  {journal} {\bibinfo  {journal} {Phys. Rev. X}\
  }\textbf {\bibinfo {volume} {12}},\ \bibinfo {pages} {041009} (\bibinfo
  {year} {2022})}\BibitemShut {NoStop}%
\bibitem [{\citenamefont {Denning}\ \emph {et~al.}(2022)\citenamefont
  {Denning}, \citenamefont {Wubs}, \citenamefont {Stenger}, \citenamefont
  {M{\o}rk},\ and\ \citenamefont {Kristensen}}]{denning2022quantum}%
  \BibitemOpen
  \bibfield  {author} {\bibinfo {author} {\bibfnamefont {E.~V.}\ \bibnamefont
  {Denning}}, \bibinfo {author} {\bibfnamefont {M.}~\bibnamefont {Wubs}},
  \bibinfo {author} {\bibfnamefont {N.}~\bibnamefont {Stenger}}, \bibinfo
  {author} {\bibfnamefont {J.}~\bibnamefont {M{\o}rk}},\ and\ \bibinfo {author}
  {\bibfnamefont {P.~T.}\ \bibnamefont {Kristensen}},\ }\bibfield  {title}
  {\bibinfo {title} {Quantum theory of two-dimensional materials coupled to
  electromagnetic resonators},\ }\href@noop {} {\bibfield  {journal} {\bibinfo
  {journal} {Phys. Rev. B}\ }\textbf {\bibinfo {volume} {105}},\ \bibinfo
  {pages} {085306} (\bibinfo {year} {2022})}\BibitemShut {NoStop}%
\bibitem [{\citenamefont {Akahane}\ \emph {et~al.}(2003)\citenamefont
  {Akahane}, \citenamefont {Asano}, \citenamefont {Song},\ and\ \citenamefont
  {Noda}}]{akahane2003high}%
  \BibitemOpen
  \bibfield  {author} {\bibinfo {author} {\bibfnamefont {Y.}~\bibnamefont
  {Akahane}}, \bibinfo {author} {\bibfnamefont {T.}~\bibnamefont {Asano}},
  \bibinfo {author} {\bibfnamefont {B.-S.}\ \bibnamefont {Song}},\ and\
  \bibinfo {author} {\bibfnamefont {S.}~\bibnamefont {Noda}},\ }\bibfield
  {title} {\bibinfo {title} {High-q photonic nanocavity in a two-dimensional
  photonic crystal},\ }\href@noop {} {\bibfield  {journal} {\bibinfo  {journal}
  {Nature}\ }\textbf {\bibinfo {volume} {425}},\ \bibinfo {pages} {944}
  (\bibinfo {year} {2003})}\BibitemShut {NoStop}%
\bibitem [{\citenamefont {Ota}\ \emph {et~al.}(2018)\citenamefont {Ota},
  \citenamefont {Katsumi}, \citenamefont {Watanabe}, \citenamefont {Iwamoto},\
  and\ \citenamefont {Arakawa}}]{ota2018topological}%
  \BibitemOpen
  \bibfield  {author} {\bibinfo {author} {\bibfnamefont {Y.}~\bibnamefont
  {Ota}}, \bibinfo {author} {\bibfnamefont {R.}~\bibnamefont {Katsumi}},
  \bibinfo {author} {\bibfnamefont {K.}~\bibnamefont {Watanabe}}, \bibinfo
  {author} {\bibfnamefont {S.}~\bibnamefont {Iwamoto}},\ and\ \bibinfo {author}
  {\bibfnamefont {Y.}~\bibnamefont {Arakawa}},\ }\bibfield  {title} {\bibinfo
  {title} {Topological photonic crystal nanocavity laser},\ }\href@noop {}
  {\bibfield  {journal} {\bibinfo  {journal} {Commun. Phys.}\ }\textbf
  {\bibinfo {volume} {1}},\ \bibinfo {pages} {86} (\bibinfo {year}
  {2018})}\BibitemShut {NoStop}%
\bibitem [{\citenamefont {Ota}\ \emph {et~al.}(2019)\citenamefont {Ota},
  \citenamefont {Liu}, \citenamefont {Katsumi}, \citenamefont {Watanabe},
  \citenamefont {Wakabayashi}, \citenamefont {Arakawa},\ and\ \citenamefont
  {Iwamoto}}]{ota2019photonic}%
  \BibitemOpen
  \bibfield  {author} {\bibinfo {author} {\bibfnamefont {Y.}~\bibnamefont
  {Ota}}, \bibinfo {author} {\bibfnamefont {F.}~\bibnamefont {Liu}}, \bibinfo
  {author} {\bibfnamefont {R.}~\bibnamefont {Katsumi}}, \bibinfo {author}
  {\bibfnamefont {K.}~\bibnamefont {Watanabe}}, \bibinfo {author}
  {\bibfnamefont {K.}~\bibnamefont {Wakabayashi}}, \bibinfo {author}
  {\bibfnamefont {Y.}~\bibnamefont {Arakawa}},\ and\ \bibinfo {author}
  {\bibfnamefont {S.}~\bibnamefont {Iwamoto}},\ }\bibfield  {title} {\bibinfo
  {title} {Photonic crystal nanocavity based on a topological corner state},\
  }\href@noop {} {\bibfield  {journal} {\bibinfo  {journal} {Optica}\ }\textbf
  {\bibinfo {volume} {6}},\ \bibinfo {pages} {786} (\bibinfo {year}
  {2019})}\BibitemShut {NoStop}%
\bibitem [{\citenamefont {Zhang}\ \emph {et~al.}(2020)\citenamefont {Zhang},
  \citenamefont {Xie}, \citenamefont {Hao}, \citenamefont {Dang}, \citenamefont
  {Xiao}, \citenamefont {Shi}, \citenamefont {Ni}, \citenamefont {Niu},
  \citenamefont {Wang}, \citenamefont {Jin} \emph {et~al.}}]{zhang2020low}%
  \BibitemOpen
  \bibfield  {author} {\bibinfo {author} {\bibfnamefont {W.}~\bibnamefont
  {Zhang}}, \bibinfo {author} {\bibfnamefont {X.}~\bibnamefont {Xie}}, \bibinfo
  {author} {\bibfnamefont {H.}~\bibnamefont {Hao}}, \bibinfo {author}
  {\bibfnamefont {J.}~\bibnamefont {Dang}}, \bibinfo {author} {\bibfnamefont
  {S.}~\bibnamefont {Xiao}}, \bibinfo {author} {\bibfnamefont {S.}~\bibnamefont
  {Shi}}, \bibinfo {author} {\bibfnamefont {H.}~\bibnamefont {Ni}}, \bibinfo
  {author} {\bibfnamefont {Z.}~\bibnamefont {Niu}}, \bibinfo {author}
  {\bibfnamefont {C.}~\bibnamefont {Wang}}, \bibinfo {author} {\bibfnamefont
  {K.}~\bibnamefont {Jin}}, \emph {et~al.},\ }\bibfield  {title} {\bibinfo
  {title} {Low-threshold topological nanolasers based on the second-order
  corner state},\ }\href@noop {} {\bibfield  {journal} {\bibinfo  {journal}
  {Light Sci. Appl.}\ }\textbf {\bibinfo {volume} {9}},\ \bibinfo {pages} {109}
  (\bibinfo {year} {2020})}\BibitemShut {NoStop}%
\bibitem [{\citenamefont {Medina-V{\'a}zquez}\ \emph
  {et~al.}(2022)\citenamefont {Medina-V{\'a}zquez}, \citenamefont
  {Murillo-Ram{\'\i}rez}, \citenamefont {Gonz{\'a}lez-Ram{\'\i}rez},\ and\
  \citenamefont {Olive-M{\'e}ndez}}]{pti_bt_2022}%
  \BibitemOpen
  \bibfield  {author} {\bibinfo {author} {\bibfnamefont {J.~A.}\ \bibnamefont
  {Medina-V{\'a}zquez}}, \bibinfo {author} {\bibfnamefont {J.~G.}\ \bibnamefont
  {Murillo-Ram{\'\i}rez}}, \bibinfo {author} {\bibfnamefont {E.~Y.}\
  \bibnamefont {Gonz{\'a}lez-Ram{\'\i}rez}},\ and\ \bibinfo {author}
  {\bibfnamefont {S.~F.}\ \bibnamefont {Olive-M{\'e}ndez}},\ }\bibfield
  {title} {\bibinfo {title} {Weak and strong coupling regimes in a topological
  photonic crystal bowtie cavity},\ }\href@noop {} {\bibfield  {journal}
  {\bibinfo  {journal} {J. Appl. Phys.}\ }\textbf {\bibinfo {volume} {132}}
  (\bibinfo {year} {2022})}\BibitemShut {NoStop}%
\bibitem [{\citenamefont {Smirnova}\ \emph {et~al.}(2020)\citenamefont
  {Smirnova}, \citenamefont {Tripathi}, \citenamefont {Kruk}, \citenamefont
  {Hwang}, \citenamefont {Kim}, \citenamefont {Park},\ and\ \citenamefont
  {Kivshar}}]{smirnova2020room}%
  \BibitemOpen
  \bibfield  {author} {\bibinfo {author} {\bibfnamefont {D.}~\bibnamefont
  {Smirnova}}, \bibinfo {author} {\bibfnamefont {A.}~\bibnamefont {Tripathi}},
  \bibinfo {author} {\bibfnamefont {S.}~\bibnamefont {Kruk}}, \bibinfo {author}
  {\bibfnamefont {M.-S.}\ \bibnamefont {Hwang}}, \bibinfo {author}
  {\bibfnamefont {H.-R.}\ \bibnamefont {Kim}}, \bibinfo {author} {\bibfnamefont
  {H.-G.}\ \bibnamefont {Park}},\ and\ \bibinfo {author} {\bibfnamefont
  {Y.}~\bibnamefont {Kivshar}},\ }\bibfield  {title} {\bibinfo {title}
  {Room-temperature lasing from nanophotonic topological cavities},\
  }\href@noop {} {\bibfield  {journal} {\bibinfo  {journal} {Light Sci. Appl.}\
  }\textbf {\bibinfo {volume} {9}},\ \bibinfo {pages} {127} (\bibinfo {year}
  {2020})}\BibitemShut {NoStop}%
\bibitem [{\citenamefont {Noh}\ \emph {et~al.}(2020)\citenamefont {Noh},
  \citenamefont {Nasari}, \citenamefont {Kim}, \citenamefont {Le-Van},
  \citenamefont {Jia}, \citenamefont {Huang},\ and\ \citenamefont
  {Kant{\'e}}}]{noh2020experimental}%
  \BibitemOpen
  \bibfield  {author} {\bibinfo {author} {\bibfnamefont {W.}~\bibnamefont
  {Noh}}, \bibinfo {author} {\bibfnamefont {H.}~\bibnamefont {Nasari}},
  \bibinfo {author} {\bibfnamefont {H.-M.}\ \bibnamefont {Kim}}, \bibinfo
  {author} {\bibfnamefont {Q.}~\bibnamefont {Le-Van}}, \bibinfo {author}
  {\bibfnamefont {Z.}~\bibnamefont {Jia}}, \bibinfo {author} {\bibfnamefont
  {C.-H.}\ \bibnamefont {Huang}},\ and\ \bibinfo {author} {\bibfnamefont
  {B.}~\bibnamefont {Kant{\'e}}},\ }\bibfield  {title} {\bibinfo {title}
  {Experimental demonstration of single-mode topological valley-hall lasing at
  telecommunication wavelength controlled by the degree of asymmetry},\
  }\href@noop {} {\bibfield  {journal} {\bibinfo  {journal} {Opt. Lett.}\
  }\textbf {\bibinfo {volume} {45}},\ \bibinfo {pages} {4108} (\bibinfo {year}
  {2020})}\BibitemShut {NoStop}%
\bibitem [{\citenamefont {Zeng}\ \emph {et~al.}(2020)\citenamefont {Zeng},
  \citenamefont {Chattopadhyay}, \citenamefont {Zhu}, \citenamefont {Qiang},
  \citenamefont {Li}, \citenamefont {Jin}, \citenamefont {Li}, \citenamefont
  {Davies}, \citenamefont {Linfield}, \citenamefont {Zhang} \emph
  {et~al.}}]{zeng2020electrically}%
  \BibitemOpen
  \bibfield  {author} {\bibinfo {author} {\bibfnamefont {Y.}~\bibnamefont
  {Zeng}}, \bibinfo {author} {\bibfnamefont {U.}~\bibnamefont {Chattopadhyay}},
  \bibinfo {author} {\bibfnamefont {B.}~\bibnamefont {Zhu}}, \bibinfo {author}
  {\bibfnamefont {B.}~\bibnamefont {Qiang}}, \bibinfo {author} {\bibfnamefont
  {J.}~\bibnamefont {Li}}, \bibinfo {author} {\bibfnamefont {Y.}~\bibnamefont
  {Jin}}, \bibinfo {author} {\bibfnamefont {L.}~\bibnamefont {Li}}, \bibinfo
  {author} {\bibfnamefont {A.~G.}\ \bibnamefont {Davies}}, \bibinfo {author}
  {\bibfnamefont {E.~H.}\ \bibnamefont {Linfield}}, \bibinfo {author}
  {\bibfnamefont {B.}~\bibnamefont {Zhang}}, \emph {et~al.},\ }\bibfield
  {title} {\bibinfo {title} {Electrically pumped topological laser with valley
  edge modes},\ }\href@noop {} {\bibfield  {journal} {\bibinfo  {journal}
  {Nature}\ }\textbf {\bibinfo {volume} {578}},\ \bibinfo {pages} {246}
  (\bibinfo {year} {2020})}\BibitemShut {NoStop}%
\bibitem [{\citenamefont {Mehrabad}\ \emph {et~al.}(2020)\citenamefont
  {Mehrabad}, \citenamefont {Foster}, \citenamefont {Dost}, \citenamefont
  {Clarke}, \citenamefont {Patil}, \citenamefont {Fox}, \citenamefont
  {Skolnick},\ and\ \citenamefont {Wilson}}]{mehrabad2020chiral}%
  \BibitemOpen
  \bibfield  {author} {\bibinfo {author} {\bibfnamefont {M.~J.}\ \bibnamefont
  {Mehrabad}}, \bibinfo {author} {\bibfnamefont {A.~P.}\ \bibnamefont
  {Foster}}, \bibinfo {author} {\bibfnamefont {R.}~\bibnamefont {Dost}},
  \bibinfo {author} {\bibfnamefont {E.}~\bibnamefont {Clarke}}, \bibinfo
  {author} {\bibfnamefont {P.~K.}\ \bibnamefont {Patil}}, \bibinfo {author}
  {\bibfnamefont {A.~M.}\ \bibnamefont {Fox}}, \bibinfo {author} {\bibfnamefont
  {M.~S.}\ \bibnamefont {Skolnick}},\ and\ \bibinfo {author} {\bibfnamefont
  {L.~R.}\ \bibnamefont {Wilson}},\ }\bibfield  {title} {\bibinfo {title}
  {Chiral topological photonics with an embedded quantum emitter},\ }\href@noop
  {} {\bibfield  {journal} {\bibinfo  {journal} {Optica}\ }\textbf {\bibinfo
  {volume} {7}},\ \bibinfo {pages} {1690} (\bibinfo {year} {2020})}\BibitemShut
  {NoStop}%
\bibitem [{\citenamefont {Gu}\ \emph {et~al.}(2021)\citenamefont {Gu},
  \citenamefont {Yuan}, \citenamefont {Zhao}, \citenamefont {Ji}, \citenamefont
  {Liu}, \citenamefont {Fang}, \citenamefont {Gan},\ and\ \citenamefont
  {Zhao}}]{gu2021topological}%
  \BibitemOpen
  \bibfield  {author} {\bibinfo {author} {\bibfnamefont {L.}~\bibnamefont
  {Gu}}, \bibinfo {author} {\bibfnamefont {Q.}~\bibnamefont {Yuan}}, \bibinfo
  {author} {\bibfnamefont {Q.}~\bibnamefont {Zhao}}, \bibinfo {author}
  {\bibfnamefont {Y.}~\bibnamefont {Ji}}, \bibinfo {author} {\bibfnamefont
  {Z.}~\bibnamefont {Liu}}, \bibinfo {author} {\bibfnamefont {L.}~\bibnamefont
  {Fang}}, \bibinfo {author} {\bibfnamefont {X.}~\bibnamefont {Gan}},\ and\
  \bibinfo {author} {\bibfnamefont {J.}~\bibnamefont {Zhao}},\ }\bibfield
  {title} {\bibinfo {title} {A topological photonic ring-resonator for on-chip
  channel filters},\ }\href@noop {} {\bibfield  {journal} {\bibinfo  {journal}
  {J. Light. Technol.}\ }\textbf {\bibinfo {volume} {39}},\ \bibinfo {pages}
  {5069} (\bibinfo {year} {2021})}\BibitemShut {NoStop}%
\bibitem [{\citenamefont {Gong}\ \emph {et~al.}(2020)\citenamefont {Gong},
  \citenamefont {Wong}, \citenamefont {Bennett}, \citenamefont {Huffaker},\
  and\ \citenamefont {Oh}}]{gong2020topological}%
  \BibitemOpen
  \bibfield  {author} {\bibinfo {author} {\bibfnamefont {Y.}~\bibnamefont
  {Gong}}, \bibinfo {author} {\bibfnamefont {S.}~\bibnamefont {Wong}}, \bibinfo
  {author} {\bibfnamefont {A.~J.}\ \bibnamefont {Bennett}}, \bibinfo {author}
  {\bibfnamefont {D.~L.}\ \bibnamefont {Huffaker}},\ and\ \bibinfo {author}
  {\bibfnamefont {S.~S.}\ \bibnamefont {Oh}},\ }\bibfield  {title} {\bibinfo
  {title} {Topological insulator laser using valley-hall photonic crystals},\
  }\href@noop {} {\bibfield  {journal} {\bibinfo  {journal} {Acs Photonics}\
  }\textbf {\bibinfo {volume} {7}},\ \bibinfo {pages} {2089} (\bibinfo {year}
  {2020})}\BibitemShut {NoStop}%
\bibitem [{\citenamefont {Liu}\ \emph {et~al.}(2022)\citenamefont {Liu},
  \citenamefont {Zhao}, \citenamefont {Zhang},\ and\ \citenamefont
  {Gao}}]{liu2022topological}%
  \BibitemOpen
  \bibfield  {author} {\bibinfo {author} {\bibfnamefont {X.}~\bibnamefont
  {Liu}}, \bibinfo {author} {\bibfnamefont {L.}~\bibnamefont {Zhao}}, \bibinfo
  {author} {\bibfnamefont {D.}~\bibnamefont {Zhang}},\ and\ \bibinfo {author}
  {\bibfnamefont {S.}~\bibnamefont {Gao}},\ }\bibfield  {title} {\bibinfo
  {title} {Topological cavity laser with valley edge states},\ }\href@noop {}
  {\bibfield  {journal} {\bibinfo  {journal} {Opt. Express}\ }\textbf {\bibinfo
  {volume} {30}},\ \bibinfo {pages} {4965} (\bibinfo {year}
  {2022})}\BibitemShut {NoStop}%
\bibitem [{\citenamefont {Song}\ \emph {et~al.}(2005)\citenamefont {Song},
  \citenamefont {Noda}, \citenamefont {Asano},\ and\ \citenamefont
  {Akahane}}]{song2005ultra}%
  \BibitemOpen
  \bibfield  {author} {\bibinfo {author} {\bibfnamefont {B.-S.}\ \bibnamefont
  {Song}}, \bibinfo {author} {\bibfnamefont {S.}~\bibnamefont {Noda}}, \bibinfo
  {author} {\bibfnamefont {T.}~\bibnamefont {Asano}},\ and\ \bibinfo {author}
  {\bibfnamefont {Y.}~\bibnamefont {Akahane}},\ }\bibfield  {title} {\bibinfo
  {title} {Ultra-high-q photonic double-heterostructure nanocavity},\
  }\href@noop {} {\bibfield  {journal} {\bibinfo  {journal} {Nat. Mater.}\
  }\textbf {\bibinfo {volume} {4}},\ \bibinfo {pages} {207} (\bibinfo {year}
  {2005})}\BibitemShut {NoStop}%
\bibitem [{\citenamefont {Charbonneau-Lefort}\ \emph
  {et~al.}(2002)\citenamefont {Charbonneau-Lefort}, \citenamefont {Istrate},
  \citenamefont {Allard}, \citenamefont {Poon},\ and\ \citenamefont
  {Sargent}}]{charbonneau2002photonic}%
  \BibitemOpen
  \bibfield  {author} {\bibinfo {author} {\bibfnamefont {M.}~\bibnamefont
  {Charbonneau-Lefort}}, \bibinfo {author} {\bibfnamefont {E.}~\bibnamefont
  {Istrate}}, \bibinfo {author} {\bibfnamefont {M.}~\bibnamefont {Allard}},
  \bibinfo {author} {\bibfnamefont {J.}~\bibnamefont {Poon}},\ and\ \bibinfo
  {author} {\bibfnamefont {E.~H.}\ \bibnamefont {Sargent}},\ }\bibfield
  {title} {\bibinfo {title} {Photonic crystal heterostructures: Waveguiding
  phenomena and methods of solution in an envelope function picture},\
  }\href@noop {} {\bibfield  {journal} {\bibinfo  {journal} {Phys. Rev. B}\
  }\textbf {\bibinfo {volume} {65}},\ \bibinfo {pages} {125318} (\bibinfo
  {year} {2002})}\BibitemShut {NoStop}%
\bibitem [{\citenamefont {Kristensen}\ \emph {et~al.}(2020)\citenamefont
  {Kristensen}, \citenamefont {Herrmann}, \citenamefont {Intravaia},\ and\
  \citenamefont {Busch}}]{kristensen2020modeling}%
  \BibitemOpen
  \bibfield  {author} {\bibinfo {author} {\bibfnamefont {P.~T.}\ \bibnamefont
  {Kristensen}}, \bibinfo {author} {\bibfnamefont {K.}~\bibnamefont
  {Herrmann}}, \bibinfo {author} {\bibfnamefont {F.}~\bibnamefont
  {Intravaia}},\ and\ \bibinfo {author} {\bibfnamefont {K.}~\bibnamefont
  {Busch}},\ }\bibfield  {title} {\bibinfo {title} {Modeling electromagnetic
  resonators using quasinormal modes},\ }\href@noop {} {\bibfield  {journal}
  {\bibinfo  {journal} {Adv. Opt. Photonics}\ }\textbf {\bibinfo {volume}
  {12}},\ \bibinfo {pages} {612} (\bibinfo {year} {2020})}\BibitemShut
  {NoStop}%
\bibitem [{\citenamefont {Christiansen}\ \emph {et~al.}(2019)\citenamefont
  {Christiansen}, \citenamefont {Wang}, \citenamefont {Sigmund},\ and\
  \citenamefont {Stobbe}}]{christiansen2019designing}%
  \BibitemOpen
  \bibfield  {author} {\bibinfo {author} {\bibfnamefont {R.~E.}\ \bibnamefont
  {Christiansen}}, \bibinfo {author} {\bibfnamefont {F.}~\bibnamefont {Wang}},
  \bibinfo {author} {\bibfnamefont {O.}~\bibnamefont {Sigmund}},\ and\ \bibinfo
  {author} {\bibfnamefont {S.}~\bibnamefont {Stobbe}},\ }\bibfield  {title}
  {\bibinfo {title} {Designing photonic topological insulators with
  quantum-spin-hall edge states using topology optimization},\ }\href@noop {}
  {\bibfield  {journal} {\bibinfo  {journal} {NanoPhoton.}\ }\textbf {\bibinfo
  {volume} {8}},\ \bibinfo {pages} {1363} (\bibinfo {year} {2019})}\BibitemShut
  {NoStop}%
\end{thebibliography}%

\end{document}